\documentclass[onecolumn]{aastex631}

\usepackage{csquotes}

\begin{document}

\title{10 Years of Archival High-Resolution NIR Spectra: The Raw and Reduced IGRINS Spectral Archive (RRISA)}

\author[0000-0002-8378-1062]{Erica Sawczynec}
\affiliation{The University of Texas at Austin, Department of Astronomy, 2515 Speedway, Stop C1400, Austin, TX 78712-1205, USA}

\author[0000-0001-6909-3856]{Kyle F. Kaplan}
\affiliation{The University of Texas at Austin, Department of Astronomy, 2515 Speedway, Stop C1400, Austin, TX 78712-1205, USA}

\author[0000-0001-7875-6391]{Gregory N. Mace}
\affiliation{The University of Texas at Austin, Department of Astronomy, 2515 Speedway, Stop C1400, Austin, TX 78712-1205, USA}

\author[0000-0003-0894-7824]{Jae-Joon Lee}
\affiliation{Korea Astronomy and Space Science Institute, 776 Daedeok-daero, Yuseong-gu, Daejeon, 34055, Republic of Korea}

\author[0000-0003-3577-3540]{Daniel T. Jaffe}
\affiliation{The University of Texas at Austin, Department of Astronomy, 2515 Speedway, Stop C1400, Austin, TX 78712-1205, USA}

\author[0000-0002-2692-7520]{Chan Park}
\affiliation{Korea Astronomy and Space Science Institute, 776 Daedeok-daero, Yuseong-gu, Daejeon, 34055, Republic of Korea}

\author{In-Soo Yuk}
\affiliation{Korea Astronomy and Space Science Institute, 776 Daedeok-daero, Yuseong-gu, Daejeon, 34055, Republic of Korea}

\author{Moo-Young Chun}
\affiliation{Korea Astronomy and Space Science Institute, 776 Daedeok-daero, Yuseong-gu, Daejeon, 34055, Republic of Korea}

\author[0000-0002-2548-238X]{Soojong Pak}
\affiliation{School of Space Research, 1 Seocheon-dong, Giheung-gu, Yongin, Gyeonggi-do, 446-701, Republic of Korea}

\author[0000-0002-2013-1273]{Narae Hwang}
\affiliation{Korea Astronomy and Space Science Institute, 776 Daedeok-daero, Yuseong-gu, Daejeon, 34055, Republic of Korea}

\author[0000-0002-4445-7366]{Ueejeong Jeong}
\affiliation{The University of Texas at Austin, Department of Astronomy, 2515 Speedway, Stop C1400, Austin, TX 78712-1205, USA}
\affiliation{Korea Astronomy and Space Science Institute, 776 Daedeok-daero, Yuseong-gu, Daejeon, 34055, Republic of Korea}

\author[0000-0003-4770-688X]{Hwihyun Kim}
\affiliation{Gemini Observatory/NSF’s NOIRLab, 950 N. Cherry Ave. Tucson, AZ 85719, USA}

\author[0000-0001-9263-3275]{Hyun-Jeong Kim}
\affiliation{Korea Astronomy and Space Science Institute, 776 Daedeok-daero, Yuseong-gu, Daejeon, 34055, Republic of Korea}

\author{Kang-Min Kim}
\affiliation{Korea Astronomy and Space Science Institute, 776 Daedeok-daero, Yuseong-gu, Daejeon, 34055, Republic of Korea}

\author{Sanghyuk Kim}
\affiliation{Korea Astronomy and Space Science Institute, 776 Daedeok-daero, Yuseong-gu, Daejeon, 34055, Republic of Korea}

\author[0000-0003-1270-9802]{Huynh Anh N. Le}
\affiliation{CAS Key Laboratory for Research in Galaxies and Cosmology, Department of Astronomy, University of Science and Technology of China, Hefei 230026, China}
\affiliation{School of Astronomy and Space Science, University of Science and technology of China, Hefei 230026, China}

\author{Hye-In Lee}
\affiliation{UJU Electronics, 3F, Daehak 3-ro, Yeongtong-gu, Suwon-si, Gyeonggi-do, 16227, Republic of Korea}

\author[0000-0001-6034-2238]{Sungho Lee}
\affiliation{Korea Astronomy and Space Science Institute, 776 Daedeok-daero, Yuseong-gu, Daejeon, 34055, Republic of Korea}

\author[0000-0002-0418-5335]{Heeyoung Oh}
\affiliation{Korea Astronomy and Space Science Institute, 776 Daedeok-daero, Yuseong-gu, Daejeon, 34055, Republic of Korea}

\author{Jae Sok Oh}
\affiliation{Korea Astronomy and Space Science Institute, 776 Daedeok-daero, Yuseong-gu, Daejeon, 34055, Republic of Korea}

\author[0000-0002-6982-7722]{Byeong-Gon Park}
\affiliation{Korea Astronomy and Space Science Institute, 776 Daedeok-daero, Yuseong-gu, Daejeon, 34055, Republic of Korea}

\author{Woojin Park}
\affiliation{Korea Astronomy and Space Science Institute, 776 Daedeok-daero, Yuseong-gu, Daejeon, 34055, Republic of Korea}

\author{Young-Sam Yu}
\affiliation{Korea Astronomy and Space Science Institute, 776 Daedeok-daero, Yuseong-gu, Daejeon, 34055, Republic of Korea}



\begin{abstract}

The Immersion GRating INfrared Spectrometer (IGRINS) is a compact, high-resolution (R$\sim$45,000) near-infrared spectrograph spanning 1.45 to 2.45 $\mu$m in a single exposure. 
We introduce the Raw and Reduced IGRINS Spectral Archive (RRISA), which provides public data access for all non-proprietary IGRINS data taken at McDonald Observatory, Lowell's Discovery Channel Telescope, and Gemini South.
RRISA provides access to raw files, reduced data products, and cross-matched IGRINS targets with the SIMBAD, 2MASS, Gaia DR3, APOGEE2 DR17, and PASTEL catalogs.
We also introduce version 3 of the IGRINS data reduction pipeline, \texttt{IGRINS PLP v3}, which implements an improved cosmic ray correction, pattern noise removal, and a new flexure correction that reduces telluric residuals.
RRISA and supporting information can be found at \url{http://igrinscontact.github.io}.

\end{abstract}

\keywords{astronomy data reduction; catalogs; near infrared astronomy; high resolution spectroscopy}


\section{Introduction} \label{sec:intro}
The Immersion GRating INfrared Spectrometer (IGRINS) \citep{Yuk2010, Park2014, Mace2016, Mace2018} employs a silicon immersion grating as the primary disperser \citep{Jaffe1998, Wang2010, Gully-Santiago2012} and two volume-phase holographic gratings as cross-dispersers, allowing simultaneous and complete coverage of both the H- and K-bands (1.45-2.45 $\mu$m) at high-resolution (R$\sim$45,000; 3.3-pixel sampling).
The instrument sits in a relatively small 90 x 60 x 40 cm cryostat that mounts at a telescope's Cassegrain focus.
IGRINS' small size and adjustable fore-optics, used to match the telescope input beams to the internal f/10 optics of the spectrometer, have allowed the instrument to observe at three telescopes since commissioning in 2014: the 2.7m Harlan J. Smith Telescope (HJST) at McDonald Observatory, the 4.3m Lowell Discovery Telescope (LDT; formally the Discovery Channel Telescope), and the 8.1m Gemini South Telescope \citep{Mace2018}.

Aside from the fore-optics, all other optics in IGRINS are fixed and produce \textit{nearly identical spectra regardless of the telescope where IGRINS is mounted}.
A reflective surface at the slit directs some of the incoming light to a K-band filtered engineering-grade slit-viewing camera that aids in target acquisition and telescope guiding (at McDonald and LDT).
IGRINS' fixed internal optics means that the slit dimensions scale inversely with telescope diameter: 1$\farcs$0 x 14$\farcs$8 on HJST, 0$\farcs$63 x 9$\farcs$3 at LDT, and 0$\farcs$32 x 4$\farcs$9 at Gemini South.
After the incoming light passes through the slit, a collimator feeds a 25 mm beam to the single silicon immersion grating that sits at an image of the telescope pupil.
The reflective optics direct the output from the immersion grating to a dichroic that splits the dispersed light into H- and K-band arms.
The arm for each band has additional optics that form white pupils on the two cross-dispersing volume-phase holographic gratings.
Finally, the cameras in each respective arm form images of the spectra on a pair of 2048 x 2048 pixel Teledyne H2RG focal plane arrays. 
The resultant raw data files from IGRINS are slit-viewing camera images and the H- and K-band echellograms.

\begin{deluxetable*}{cc}\label{tab:travel}

    \centering
    \tablecaption{IGRINS visiting schedule for each telescope (YYYYMMDD).}
    \tablehead{
    \colhead{Telescope} & \colhead{Dates Visited}} 
    \startdata
    HJST & 20140707 -- 20160727 \\
    LDT & 20160908 -- 20170222 \\
    HJST & 20170310 -- 20170813 \\
    LDT & 20170829 -- 20180124 \\
    Gemini South & 20180401 -- 20180706 \\
    LDT & 20180915 -- 20190430 \\
    Gemini South & 20200130 -- 20240421 \\
    HJST & 20241111   --     Present     \\
    \enddata
\end{deluxetable*}

IGRINS was commissioned in 2014 on the HJST and has followed the travel schedule outlined in Table \ref{tab:travel}.
IGRINS saw $\sim 1$ magnitude deeper at LDT and $\sim 2.5$ magnitudes deeper at Gemini South, compared to HJST \citep{Mace2018}.
Table \ref{tab:exp-time} outlines IGRINS exposure times per frame, given average clear observing conditions at each telescope, required to reach a signal-to-noise ratio (SNR) in K-band of 100 per resolution element.
An updated copy of IGRINS, named IGRINS-2, is being commissioned at Gemini North as a facility instrument and will come on-sky for science operations in 2025 \citep{Oh2024}.


\begin{deluxetable*}{cccc}\label{tab:exp-time}
    \centering
    \tablecaption{Exposure time estimates in seconds per frame to reach a signal-to-noise ratio of 100 per resolution element. These assumptions are for typical observing conditions at each telescope and a single ABBA nod sequence of four exposures unless prefix multipliers (2x or 4x) are given. We assume seeing of 1 and 0.8 arcseconds respectively for McDonald and LDT. For Gemini South, we assume conditions corresponding to an image quality of 70, cloud cover of 70\%, sky background of any kind, and water vapor of any percentage which are standard observing conditions for most Gemini South IGRINS observations.}
    \tablehead{
        \colhead{K-Band Mangitude} & \colhead{McDonald} & \colhead{LDT} & \colhead{GS}} 
    \startdata
    4 & $< 20$    & $< 15$   & -- \\
    5 & $< 20$    & $< 15$   & 1.63 \\
    6 & 30        & $< 15$   & 4 \\
    7 & 80        & 20       & 10 \\
    8 & 190       & 50       & 25 \\
    9 & 2x 300  & 130      & 65 \\
    10 & 4x 300 & 2x 300 & 160 \\
    11 & --       & 4x 300 & 2x 300 \\
    12 & --       & --       & 4x 300 \\
    13 & --       & --       & --\tablenotemark{a} \\
    \enddata
    \tablenotetext{a}{Objects fainter than K=12 mag are not observable in the average Gemini South conditions assumed. In better conditions (IQ-20, CC-50\%), objects as faint as K=15 mag are possible.}
\end{deluxetable*}

The science applications of IGRINS are broad and include the study of:
exoplanet atmospheres \citep{Flagg2019, Line2021,Ridden-Harper2023, Brogi2023, Smith2024};
Neptune's moon Triton \citep{Tegler2019};
a spectral atlas of a T6 brown dwarf \citep{Tannock2022};
the IGRINS spectral library \citep{Park2018};
stellar parameters \citep{Afsar2016, afsar2018, lopezvaldivia2019, afsar2023, Brady2023, Holanda2024, Lim2024, Nandakumar2024};
exoplanet validation \citep{Johns-Krull2016, Mann2016b, Mann2017, Rizzuto2017, Mann2018};
multiple star systems \citep{Rappaport2017, Mace2018sci,Tang2023};
planetary nebulae and the interstellar medium \citep{Oh2016b, Sterling2016, Oh2018, Kaplan2017, Lacy2017, Le2017, Kaplan2021};
stellar magnetic fields \citep{Han2017, Kesseli2018, Han2019, Han2023, Sokal2020};
and young stellar objects (YSOs) \citep{lee2016, Oh2016, gully2017, lyo2017, Sokal2018, Park2020, Park2021, Kidder2021, lopezvaldivia2021, lopezvaldivia2023, wang2023, tang2024}.
Some of the IGRINS YSO spectra are also available in the Spectroscopy of Exoplanet-forming Disks \href{https://www.spexodisks.com}{(SpExoDisks)} database.

As of May 2023, IGRINS has completed over 17,700 observations of science targets and telluric standards. 
Only a small fraction of those data have been accessible to the astronomy community since there has not been a centralized public database for IGRINS.
We introduce the Raw and Reduced IGRINS Spectral Archive (RRISA), which streamlines public access to IGRINS data products.
RRISA currently features IGRINS data taken between August 2014 and May 2023 and encompasses more than 1,000 nights on sky, resulting in almost 3,500 unique target observations from over 4,000 hours of total observing time.
Almost all science target observations are paired with a nearby telluric standard star for relative flux calibration, and RRISA includes those $>$790 unique standard stars observations.
For RRISA ease of use, we have linked IGRINS targets to SIMBAD \citep{Wenger2000} identifiers whenever possible and cross-matched with the 2MASS \citep{Skrutskie2006}, Gaia DR3 \citep{Gaia2016, Gaia2023} APOGEE2 DR 17 \citep{Abdurrouf2022}, and PASTEL \citep{Soubiran2016} catalogs to provide identifiers and additional survey-specific data for IGRINS targets.
 
In Section \ref{sec:org}, we outline the steps to creating RRISA: IGRINS raw and reduced data logging, target identification, cross-matching, and public access. 
In Section \ref{sec:PLP}, we outline the IGRINS data reduction process
and highlight the recent improvements to the IGRINS Pipeline Package--used to reduce the data featured in RRISA--including cosmic ray removal, pattern noise removal, and flexure correction.
Finally, in Section \ref{sec:tools}, we discuss IGRINS-compatible Python data analysis tools and publicly available RRISA tutorials.

\section{IGRINS Data Logging}\label{sec:org}
There are raw and reduced data organization codes that log every IGRINS raw spectrum and every spectrum reduced using \texttt{IGRINS PLP} v3 (see Section \ref{sec:PLP}).
The raw data organization code scrapes information from the headers of each H-band raw image and the IGRINS digital night logs (report information about every raw frame taken on a night of observing). 
IGRINS takes simultaneous H- and K-band spectra, so the digital logs are generated using only the H-band files since the information will be nearly identical between the two bands.
The reduced data organization code builds on the the raw log with information from the \texttt{IGRINS PLP} v3 recipe logs which indicate which raw frames are used in each reduction.
The \texttt{IGRINS PLP} v3 does not reduce all raw IGRINS frames, as some are bad quality or require a specialty reduction (e.g. when there are multiple objects on the slit or extended sources).
The information we gather for each raw frame and reduced spectrum makes up the bulk of the RRISA Raw and Reduced components and their content is explicitly stated in Table \ref{tab:content_summary} marked with the check marks in the Raw and Reduced columns.



\subsection{Target Identification}
We start IGRINS target identification by searching the recorded target names for each observation using SIMBAD, but since there was no requirement at any telescope IGRINS visited for target names to be complete or common, we find that $\sim 20\%$ of the object names in RRISA's reduced log (``OBJNAME\_super'' in Table \ref{tab:content_summary}) were not found in SIMBAD.
We elected to fix unsearchable names by hand to ensure the correct identification of as many objects as possible.
We started by searching within five arcminutes of the reported target coordinates to see if any of the object names in SIMBAD were similar to the recorded target name.
In most cases, this was successful in identifying the SIMBAD searchable target name.
Most of the automated name search failures arose due to a departure from strict SIMBAD nomenclature requirements (e.g. KPNO-Tau 15 instead of the SIMBAD name [BLH2002] KPNO-Tau 15).
In the remaining cases, we referred to the IGRINS paper observation logs and searched the observer's written coordinates and/or object name.
We were able to confidently identify most of these objects, leaving only 828 of the original 17,730 ($\sim 4.7\%$) reduced objects unidentified.
However, that number includes objects that have correct names but are not SIMBAD searchable (213 of the 828 or $\sim 26\%$ of non searchable objects).
The remaining 615 unidentified objects ($\sim 3.4\%$ of the total reduced IGRINS sample) are distributed per telescope as follows: 146 at McDonald, 137 at LDT, and 332 at Gemini South.
When searching SIMBAD with the coordinates for these unidentified targets, there are no results returned with names similar to those recorded in the IGRINS logs, so we cannot identify them with any confidence.
Comparing the target names to the program information at Gemini, it is clear that many of these objects are not in the SIMBAD catalog and are unpublished stellar cluster members.

We elect to use hand correction to identify IGRINS targets with unsearchable names over other automated target identifying methods like using coordinate searches, observation documentation, or slit camera images due to several quirks unique to IGRINS observations.
The coordinates recorded in IGRINS headers mark a reference pixel off to the side of the IGRINS slit, not the location of the target, and the reference pixel location can be changed at any time by the observer without note\footnote{See the slit images in Section 4 of the \href{https://cloud.wikis.utexas.edu/wiki/spaces/IGRINS/pages/78907201/Using+the+IGRINS+Data+Taking+Software}{IGRINS Data Taking Software Tutorial} on the \href{https://cloud.wikis.utexas.edu/wiki/spaces/IGRINS/overview?homepageId=78905352}{IGRINS Wiki}. The green pixel highlighted in the images off to the left of the slit is the standard coordinate reference pixel.}.
Additionally, at HJST, the header coordinates are a function of the user-calibrated telescope pointing and can be several arcminutes off from the true object coordinates, making it difficult to identify objects by coordinate searches alone, especially in crowded fields.
The original IGRINS proposals and any corresponding finder charts were not archived for HJST or LDT and are not available for Gemini South. 
Observations at both HJST and LDT were documented using paper logs that frequently have missing, incomplete, or the same unsearchable target names and/or coordinates as the digital night logs and the IGRINS headers.


IGRINS uses a K-band engineering-grade slit-viewing camera to verify target acquisition on-sky.
However, the FOV for the camera is small, reaching 93 arcseconds west of the slit, 19 arcseconds east of the slit, 87 arcseconds north of the slit, and 100 arcseconds south of the slit\footnote{\href{https://cloud.wikis.utexas.edu/wiki/spaces/IGRINS/pages/78907172/Slit+Viewer+Field-of-View}{At HJST} where the slit is largest on-sky.}.
Since the field of view for the slit camera is small and the exposure times on most images are short\footnote{The slit camera images are used for guiding the telescope during spectra acquisition, and faster exposure times result in better guiding on the slit.}, the only object usually visible in the slit camera image is the target.
Apart from observations at Gemini South, saving slit camera images has always been up to the observer's discretion since saving many slit camera images can rapidly fill the IGRINS hard drives, meaning there are many nights with no slit camera images available.
In RRISA Raw, the reference slit camera image file numbers that are linked to specific raw spectra files are provided (``SDCS'' in Table \ref{tab:content_summary}).
We also generated a separate log of every slit camera image file IGRINS has ever saved, so users interested in reviewing all slit-viewing images can do so.
The slit camera images are available to download via the RRISA Raw data folders, when available, or through the log of saved slit camera image files (See Section \ref{sec:access}).



\startlongtable
\begin{deluxetable*}{cp{3in}ccc}\label{tab:content_summary}
\centering
\tablecaption{An outline of the information that is included in each component of RRISA with detailed descriptions for each field.}
\tabletypesize{\scriptsize}
\tablehead{
\colhead{Name} & \colhead{Description} & 
\colhead{Raw} & \colhead{Reduced} & 
\colhead{XMatch}} 
\startdata
OBJNAME/OBJNAME\_super & The object name in the virtual observing logs & \checkmark & \checkmark & \checkmark \\
NAME & The hand-corrected object name & \nodata & \checkmark & \checkmark \\
OBJNAME\_recipe & The object name in the \texttt{IGRINS PLP v3} recipe log & \nodata & \checkmark & \checkmark \\
MAIN\_ID & The primary SIMBAD identifier & \nodata & \nodata & \checkmark \\
IDS & Object name aliases (from SIMBAD) & \nodata & \nodata & \checkmark \\
for RA & Right ascension recorded by the telescope (in J2000)& \checkmark & \checkmark & \checkmark \\
DEC & Declination recorded by the telescope (in J2000) & \checkmark & \checkmark & \checkmark \\
RA\_s & Right ascention from SIMBAD & \nodata & \nodata & \checkmark \\
DEC\_s & Declination from SIMBAD & \nodata & \nodata & \checkmark \\
OBJTYPE & IGRINS frame type (STD, TAR, SKY, FLAT, DARK, or ARC) & \checkmark & \checkmark & \checkmark \\
FILENAME & IGRINS filename &\checkmark & \checkmark & \checkmark \\
FILENUMBER & IGRINS file number & \checkmark & \checkmark & \checkmark \\
FRAMETYPE & A/B (nod on the slit) or ON/OFF (usually only for flats to describe lamp state) & \checkmark & \nodata & \nodata \\
CIVIL & Local date of observation (YYYYMMDD) & \checkmark & \checkmark & \checkmark \\
JD & Julian date when observation finished & \checkmark & \checkmark & \checkmark \\
OBSTIME & Time when observation began (HH:MM:SS; universal time) & \checkmark & \checkmark & \checkmark \\
EXPTIME & Exposure time (in seconds) & \checkmark & \checkmark & \checkmark \\
ROTPA & Position angle of the slit (deg E of N) & \checkmark & \checkmark & \checkmark \\
AM & Airmass at the end of observation & \checkmark & \checkmark & \checkmark \\
BVC & Barycentric velocity correction (in km/s) & \checkmark & \checkmark & \checkmark \\
FACILITY & Telescope of observation (McDonald, DCT, or Gemini South) & \checkmark & \checkmark & \checkmark \\
PI & Principal investigator for observations (only available for McDonald and DCT) & \checkmark & \checkmark & \checkmark \\
PROGID & Gemini South program ID & \checkmark & \checkmark & \checkmark \\
FILES & List of raw file numbers used in data reduction & \nodata & \checkmark & \checkmark \\
SNRH\_pix\tablenotemark{\scriptsize a} & SNR in H-band (per pixel) & \nodata & \checkmark & \checkmark \\
SNRH\_res\tablenotemark{\scriptsize b} & SNR in H-band (per resolution element) & \nodata & \checkmark & \checkmark \\
SNRK\_pix\tablenotemark{\scriptsize a} & SNR in K-band (per pixel) & \nodata & \checkmark & \checkmark \\
SNRK\_res\tablenotemark{\scriptsize b} & SNR in K-band (per resolution element) & \nodata & \checkmark & \checkmark \\
SDCS & Either a flag followed by the number of SDCS images taken throughout the night (e.g. "-1 42") if the night has unlinked slit camera image files or a list of the slit camera image file numbers that were taken during the acquisition of the .spec file separated by a space (e.g. 100 101 102). & \checkmark & \nodata & \nodata \\
RAW\_URL & Link to raw data folder on Box & \checkmark & \checkmark & \checkmark \\
FILE\_URL & Link to download \texttt{IGRINS PLP} v3 reduced H- and K-band data products on Box & \nodata & \checkmark & \checkmark \\
CAL\_URL & Link to download associated H- and K-band calibration files for the night of observations & \nodata & \checkmark & \checkmark \\
OTYPE & SIMBAD \href{https://simbad.cds.unistra.fr/guide/otypes.htx}{object type} & \nodata & \nodata & \checkmark \\
SP\_TYPE   & Spectral type in SIMBAD & \nodata & \nodata & \checkmark \\
SP\_BIBCODE   & NASA ADS bibcode for spectral type reference & \nodata & \nodata & \checkmark \\
PMRA & RA proper motion from SIMBAD (in mas/yr) & \nodata & \nodata & \checkmark \\
PMDEC & DEC proper motion from SIMBAD (in mas/yr) & \nodata & \nodata & \checkmark \\
PM\_BIBCODE & NASA ADS bibcode for proper motion reference & \nodata & \nodata & \checkmark \\
RV\_VALUE & Radial velocity from SIMBAD (in km/s) & \nodata & \nodata & \checkmark \\
RV\_BIBCODE & NASA ADS bibcode for radial velocity reference & \nodata & \nodata & \checkmark \\
PLX\_VALUE & Parallax value from SIMBAD (in mas) & \nodata & \nodata & \checkmark \\
PLX\_BIBCODE & NASA ADS bibcode for parallax value reference & \nodata & \nodata & \checkmark \\
U,B,V,R,G,I,J,H,K & Magnitude values from SIMBAD & \nodata & \nodata & \checkmark \\
2MASS\_ID & 2MASS identifier for object & \nodata & \nodata & \checkmark \\
2MASS\_J, H, K & 2MASS J-, H-, and K- band magnitudes & \nodata & \nodata & \checkmark \\
2MASS\_Flag & Indicates if user should independently verify the 2MASS cross-match (if 1) & \nodata & \nodata & \checkmark \\
GaiaDR3\_source & Gaia DR3 identifier & \nodata & \nodata & \checkmark \\
GaiaDR3\_parallax & Gaia DR3 parallax (in mas) & \nodata & \nodata & \checkmark \\
GaiaDR3\_pm & Gaia DR3 proper motion (in mas/yr) & \nodata & \nodata & \checkmark \\
GaiaDR3\_bprp & Differential Gaia DR3 photometry: G$_{\textrm{BP}}$ - G$_{\textrm{RP}}$ & \nodata & \nodata & \checkmark \\
GaiaDR3\_ebprp & Gaia DR3 G$_{\textrm{BP}}$ - G$_{\textrm{RP}}$ excess factor & \nodata & \nodata & \checkmark \\
GaiaDR3\_gmag & Gaia DR3 G-band mean magnitude & \nodata & \nodata & \checkmark \\
GaiaDR3\_RV & Gaia DR3 radial velocity (in km/s) & \nodata & \nodata & \checkmark \\
GaiaDR3\_teff\tablenotemark{\scriptsize c} & Gaia DR3 effective temperature (in K) & \nodata & \nodata & \checkmark \\
GaiaDR3\_logg\tablenotemark{\scriptsize c} & Gaia DR3 surface gravity (in $\log_{10}(\textrm{cm/s}^2)$) & \nodata & \nodata & \checkmark \\
GaiaDR3\_FeH\tablenotemark{\scriptsize c} & Gaia DR3 metallicity (in dex) & \nodata & \nodata & \checkmark \\
GaiaDR3\_dist\tablenotemark{\scriptsize c} & Gaia DR3 distance (in pc) & \nodata & \nodata & \checkmark \\
GaiaDR3\_Flag & Indicates if user should independently verify the Gaia DR3 cross-match (if 1) & \nodata & \nodata & \checkmark \\
APOGEE2\_HRV & APOGEE2 DR17 radial velocity (in km/s) & \nodata & \nodata & \checkmark \\
APOGEE2\_teff & APOGEE2 DR17 effective temperature (in K) & \nodata & \nodata & \checkmark \\
APOGEE2\_logg & APOGEE2 DR17 surface gravity (in $\log_{10}(\textrm{cm/s}^2)$) & \nodata & \nodata & \checkmark \\
APOGEE2\_Vsini & APOGEE2 DR17 V$\sin(\textrm{i})$ (in km/s) & \nodata & \nodata & \checkmark \\
APOGEE2\_$[$M\/H$]$ & APOGEE2 DR17 metals/hydrogen (in dex) & \nodata & \nodata & \checkmark \\
APOGEE2\_$[$a\/H$]$ & APOGEE2 DR17 alpha elements/hydrogen (in dex) & \nodata & \nodata & \checkmark \\
APOGEE2\_$[$Fe\/H$]$ & APOGEE2 DR17 metallicity (in dex) & \nodata & \nodata & \checkmark \\
PASTEL\_Teff & PASTEL effective temperature value (in K) & \nodata & \nodata & \checkmark \\
PASTEL\_logg & PASTEL surface gravity value (in $\log_{10}(\textrm{cm/s}^2)$) & \nodata & \nodata & \checkmark \\
PASTEL\_$[$Fe\/H$]$ & PASTEL metallicity value (in dex) & \nodata & \nodata & \checkmark \\
PASTEL\_Flag & Indicates if user should independently verify the PASTEL cross-match (if 1) & \nodata & \nodata & \checkmark \\
\enddata
\tablenotetext{a}{Calculated by dividing the flux by the square root of the variance per pixel.}
\tablenotetext{b}{Estimated by multiplying the SNR per pixel by the square root of the amount of pixels in each resolution element (wavelength dependent).}
\tablenotetext{c}{Property derived using General Stellar Parametrizer from Gaia DR3 Phototometry (\href{https://gea.esac.esa.int/archive/documentation/GDR3/Data_analysis/chap_cu8par/sec_cu8par_apsis/ssec_cu8par_apsis_gspphot.html}{GSP-Phot}).}
\end{deluxetable*}

\subsection{Target Cross-Matching}\label{subsec:xmatch}
For the XMatch component of RRISA, we selected five catalogs we thought contained information that would complement IGRINS spectra: SIMBAD \citep{Wenger2000}, 2MASS \citep{Cutri2003}, Gaia DR3 \citep{Gaia2016, Gaia2023}, APOGEE2 DR17 \citep{Abdurrouf2022}, and PASTEL \citep{Soubiran2016}.
SIMBAD includes an extensive amount of astrophysical parameters for well-documented objects. 
Most importantly for RRISA, SIMBAD lists known aliases for each object.
2MASS provides NIR photometry for J-, H-, and K-bands.
Gaia DR3 catalogs astrometric solutions and provides astrophysical parameters derived from low-resolution (R$\sim 50-160$) BP/RP optical spectra.
APOGEE2 DR17 includes abundance measurements and astrophysical parameters derived from mid-resolution (R $\sim 22,500$) H-band spectra.
PASTEL collects astrophysical parameters derived from mid-resolution (R $\geq$ 25,000) medium SNR (S/N $\geq$ 50) spectra published by independent literature sources with smaller sample sizes that can provide higher quality astrophysical parameters than the other large catalogs included in RRISA XMatch.

We started by searching the hand-corrected names from RRISA Reduced in SIMBAD using \texttt{astroquery.simbad.query\_tap}\footnote{\href{https://www.ivoa.net/documents/TAP/}{Table Access Protocol (TAP)} defines a standardized method for accessing, searching, and returning information available in table-based databases, like astronomical catalogs, using \href{https://www.ivoa.net/documents/ADQL/2.0}{Astronomical Data Query Language} (ADQL; similar to Structured Query Language, SQL).}.
The coordinates from IGRINS logs can be offset from the true coordinates of an object, so we elected to use an object's SIMBAD coordinates for cross-matching with other catalogs.
This means that objects that do not have SIMBAD identifiers are not cross-matched and are not included in RRISA XMatch.

We cross-identify IGRINS targets with 2MASS, Gaia DR3, APOGEE2 DR17, and PASTEL using \texttt{astroquery.xmatch.query}, which allows users to search for objects in any Vizier \citep{Ochsenbein2000} catalog using coordinates.
Using the RA and DEC from SIMBAD for each IGRINS target, we searched for any catalog matches within 20 arcseconds and retained only the closest coordinate match from the catalog.
To verify that the cross-match is correct, we use \texttt{astroquery.simbad.query\_tap} to find the SIMBAD primary identifier for the catalog identifier.
We then check that the SIMBAD primary identifier already linked to the IGRINS target matches the SIMBAD primary identifier returned for the catalog identifier.
In cases where the SIMBAD primary identifiers did not match, the cross-match was deemed a mismatch and the catalog information was dropped for that IGRINS target.
While SIMBAD includes most identifiers from large catalogs, we found that there are missing aliases; if a catalog identifier is not SIMBAD searchable, the cross-match is flagged so users of RRISA can verify the integrity of the cross-match independently.
The specific IGRINS and catalog information included in the RRISA XMatch component are described in Table \ref{tab:content_summary}, any field with a check mark in the XMatch column.

\begin{deluxetable*}{cccc|c}\label{tab:tele-stat}
    \centering
    \tablecaption{IGRINS observing statistics divided by telescope for objects observed between 20140707 and 20230507.}
    \tablehead{
        \colhead{Stat} & \colhead{HJST} & \colhead{LDT} & \colhead{GS} & \colhead{Total}} 
    \startdata
    Nights on Sky & 361 & 289 & 443 & 1,093 \\
    Observing Hours & 1,632 & 1,345 & 1,086 & 4,063 \\
    Raw Frames & 56,278 & 34,201 & 51,474 & 141,953 \\
    Reduced Spectra & 8,214 & 8,890 & 18,356 & 35,460 \\
    Unique Targets\tablenotemark{\scriptsize a} & 1,385 & 807 & 1,240 & 3,432 \\
    Unique Standards\tablenotemark{\scriptsize a} & 336 & 70 & 386 & 792 \\ \hline
    \enddata
    \tablenotetext{a}{Unique to the IGRINS total sample, not the specific telescope. Object names come from SIMBAD, so any part of the IGRINS sample not in the SIMBAD catalog is not counted here--counts provide a lower limit of unique objects IGRINS has observed.}
\end{deluxetable*}

\subsection{The IGRINS Sample}

In the past ten years, IGRINS has spent 1,093 nights on-sky, which is roughly equivalent to three consecutive years of observing every night.
On those nights, IGRINS spent over 4,000 hours actively taking data, which is roughly six full months of continuous acquisition time for targets and standard stars. 
Including calibrations, IGRINS has taken around 140,000 raw spectral frames, used to produce 35,000 H- and K-band reduced spectra of over 4,000 unique targets and standard stars.
These numbers are broken down on a per-telescope basis in Table \ref{tab:tele-stat} where we can see IGRINS has spent the most time on-sky and produced the most reduced spectra while visiting Gemini South but observed the most raw frames and unique targets at HJST.

\begin{figure}
    \centering
    \includegraphics[width=0.75\linewidth]{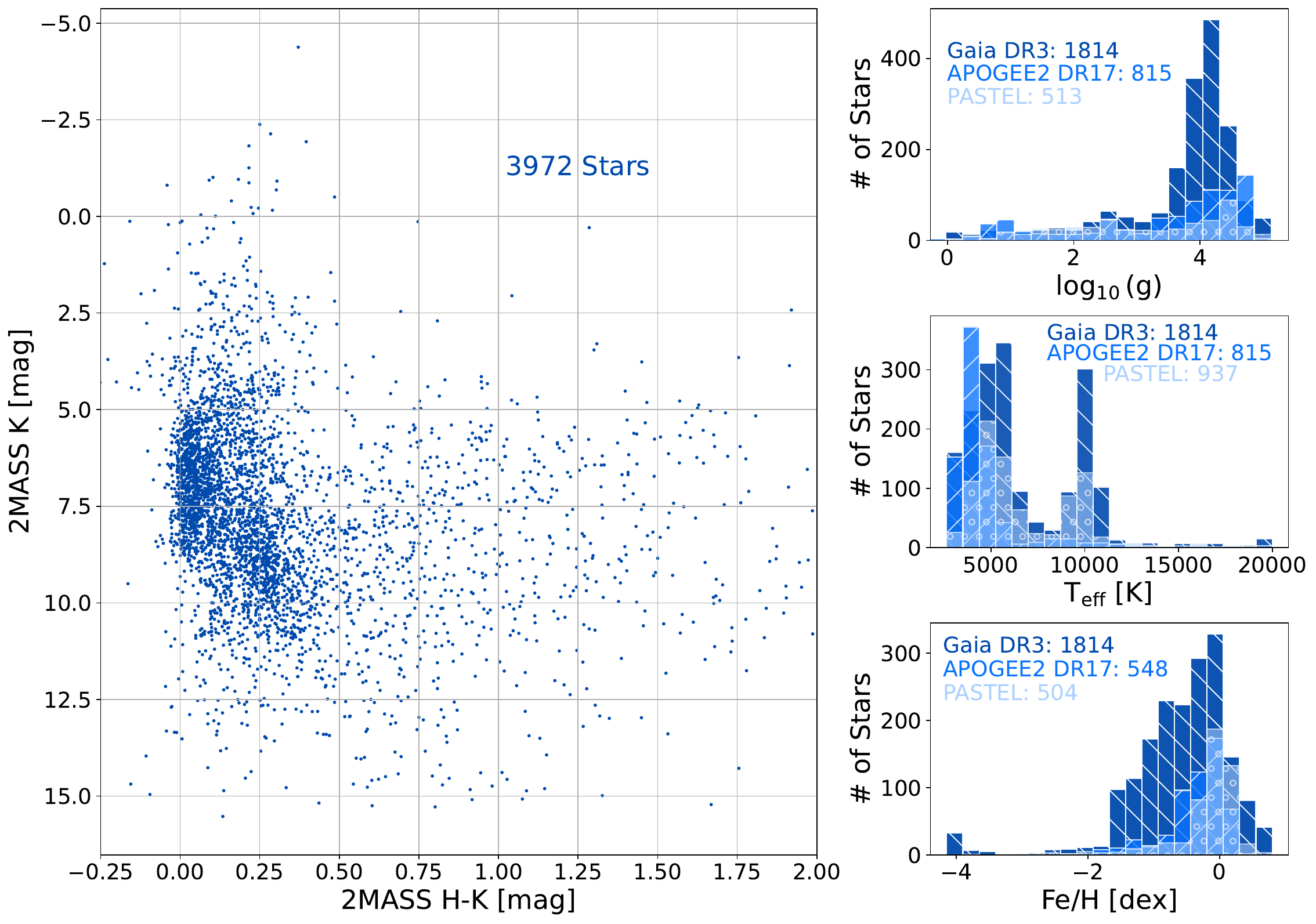}
    \caption{\emph{Left:} The 2MASS H-K color-magnitude diagram of the unique IGRINS cross-matched sample. \emph{Right:} From top to bottom, histograms of $\log_{10}(\textrm{g})$, T$_{\textrm{eff}}$, and $[$Fe/H$]$ showing the distribution of each parameter for the IGRINS targets cross-matched with Gaia DR3 (dark blue, back slashed), APOGEE2 DR17 (medium blue, forward slashed), and PASTEL (light blue, circled). \emph{Note: The IGRINS cross-matched sample includes standard stars with T$_{\textrm{eff}}$ $\sim$10,000K.}}
    \label{fig:xmatch_stats}
\end{figure}

Since IGRINS logs do not contain any target demographic information, all of our information about the IGRINS sample observed over the past ten years comes from the information in RRISA XMatch.
Figure \ref{fig:xmatch_stats} shows some of the information available in RRISA XMatch for objects in the IGRINS sample made possible by cross-matching with 2MASS, Gaia DR3, APOGEE2 DR17, and PASTEL.
Additionally, Table \ref{tab:SIMBAD_type} breaks down the most popular types of objects in the IGRINS sample using SIMBAD's object categorization--IGRINS predominantly observes different classifications of stars.

\begin{deluxetable*}{cccc|c}\label{tab:SIMBAD_type}
    \centering
    \tablecaption{All of the SIMBAD object types featured in RRISA XMatch that have \emph{over} 50 unique objects observed.}
    \tablehead{
        \colhead{Object Description} & \colhead{SIMBAD Object Type} & \colhead{RRISA Count}} 
    \startdata
        Star & * & 931 \\
        High Proper Motion Star & PM* & 677 \\
        Young Stellar Object & Y*O & 319 \\
        T Tauri Star & TT* & 223 \\
        Orion Variable & Or* & 193 \\
        Spectroscopic Binary & SB* & 167 \\
        Double or Multiple Star & ** & 161 \\
        Eruptive Variable & Er* & 115 \\
        Red Giant Branch Star & RG* & 86 \\
        Low Mass Star & LM* & 85 \\
        Brown Dwarf & BD* & 79 \\
        Herbig Ae/Be Star & Ae* & 75 \\
        Red Super Giant Star & s*r & 72 \\
        RS CVn Variable & RS* & 56 \\
        BY Dra Variable & BY* & 53 \\
    \enddata
\end{deluxetable*}

\subsection{Data Packaging \& Access}\label{sec:access}
All of the raw and reduced data from IGRINS cataloged in the three components of RRISA are hosted on the cloud-based file-sharing platform \href{https://www.box.com/home}{Box}.
Raw data can be downloaded per night through a browser using the unique access links included in the Raw, Reduced, and XMatch components of RRISA.
We package the most relevant H- and K-band reduced data files, outlined in detail in Table \ref{tab:reduced_files}, into a single compressed file for each target and standard star observed.
The reduced calibration files for a night, see Table \ref{tab:reduced_files}, are also packaged into a separate compressed file.
The compressed target and calibration files are organized by night of observation and available to download individually using the links in the Reduced and XMatch components of RRISA.
Additional ancillary files of interest such as the IGRINS hand-written observing logs and the recipe logs used to reduce IGRINS data can be downloaded individually using the \href{https://utexas.box.com/s/l7i5dbtss084mhxywmc6r8pl39bs1yhw}{RRISA Box folder}, but no download links for these files are provided in RRISA directly.

Each RRISA component can be downloaded via \href{https://github.com/IGRINScontact/RRISA}{GitHub}.
There is a folder for each RRISA component on GitHub, and each folder contains a markdown file outlining the contents included in that particular RRISA component.
Users can learn more about the contents of RRISA by visiting the RRISA website: \href{https://igrinscontact.github.io/}{igrinscontact.github.io}.

\startlongtable
\begin{deluxetable*}{cp{1.5in}p{3in}cc}\label{tab:reduced_files}
\centering
\tablecaption{Contents the compressed files for H- and K-band from the reduced data products (available per file number) and for the calibration files (available per night).}
\tabletypesize{\scriptsize}
\tablehead{
\colhead{File Extension} & \colhead{File Description} & \colhead{Notes} & \colhead{Target} & \colhead{Standard}} 
\startdata
\multicolumn{5}{c}{Included in Reduced Data Compressed Files} \\ \hline
.spec$\_$a0v.fits & Contains extensions that store different aspects of the reduced 1D spectra & Extension descriptions: 
                     \begin{itemize}
                        \item[--] $[0]$ Primary: Only contains the header with the information about the observation
                        \item[--] $[1]$ SPEC\_DIVIDE\_A0V: The corrected target spectrum = ($[4]$/$[6]$)$\cdot [7]$. This spectrum is roughly telluric corrected and relatively flux calibrated.
                        \item[--] $[2]$ SPEC\_DIVIDE\_A0V\_VARIANCE: The variance of $[1]$ (per pixel) propagated from the variance in the individual target and standard spectrum.
                        \item[--] $[3]$ WAVELENGTH: The wavelength solution from the OH sky lines (in um).
                        \item[--] $[4]$ TGT\_SPEC: The extracted target spectrum (same as in spec.fits file).
                        \item[--] $[5]$ TGT\_SPEC\_VARIANCE: The variance of $[4]$ (per pixel).
                        \item[--] $[6]$ A0V\_SPEC: The extracted A0V spectrum used for the standard star division. The observation ID (file number) for the A0V used is available as the ‘OBSID’ keyword in the header for this extension.
                        \item[--] $[7]$ A0V\_SPEC\_VARIANCE: The variance of $[6]$ (per pixel).
                        \item[--] $[8]$ VEGA\_SPEC: A synthetic model of the Vega spectrum.
                        \item[--] $[9]$ SPEC\_DIVIDE\_CONT: Similar to $[1]$, but the target spectrum is divided by the estimated continuum of the A0V star instead of the A0V spectrum itself. The resulting spectrum is relatively flux calibrated but not telluric corrected.
                        \item[--] $[10]$ SPEC\_DIVIDE\_CONT\_VARIANCE: The variance of $[9]$ (per pixel).
                        \item[--] $[11]$ MASK: The pixels used in the A0V star continuum fit for $[9]$. A value of 1 is not used in the continuum fit.
                     \end{itemize} & \checkmark & \nodata \\
.spec.fits & The extracted 1D spectra. & Columns are x-pixel position on the detector and rows are echelle order. The pixel values are the extracted counts.  & \checkmark & \checkmark \\
.sn.fits & The signal-to-noise (per resolution element) for the extracted 1D spectrum. & We recommend using .variance.fits for uncertainty. Columns are x-pixel position on the detector and rows are echelle order. & \checkmark & \checkmark \\
.variance.fits & Variance for the 1D spectra in spec.fits. & Columns are x-pixel position on the detector and rows are echelle order. The pixel values are the extracted counts. & \checkmark & \checkmark \\
.spec2d.fits & The 2D spectrum stored as a data-cube of the rectified echelle orders. & The x-axis is pixel across the detector and y-axis is the spatial axis along the IGRINS slit. The pixel values are detector counts. & \checkmark & \checkmark \\
.var2d.fits & The variance (per pixel) of the 2D spectrum. & The x-axis is pixel across the detector and y-axis is the spatial axis along the IGRINS slit.  & & \\
.slit$\_$profile.json & The average measured slit profile for a target AB nodded on the slit. & The array labeled ``profile\_x'' is the fractional distance across the slit and ``profile\_y'' is the slit profile of the target. & \checkmark & \checkmark \\
.spec\_flattened.fits & The standard star spectrum divided by an estimate of its own continuum. & Columns are x-pixel position on the detector and rows are echelle order. & \nodata & \checkmark \\
.1d$\_$plots.pdf & Extension $[1]$ of spec.a0v.fits plotted by order. & The y-axis limits of each order are set using the 99th flux percentile of the trimmed order, but usually this is not ideal for anything other than quick-looking the quality of the telluric correction. & \checkmark &  \nodata \\ \hline
\multicolumn{5}{c}{Included in Night Calibration Compressed Files} \\ \hline
.flat\_off.fits & Stack of flat OFF exposures. &  & & \\
.flat\_on.fits & Stack of flat ON exposures. &  & & \\
.flat\_normed.fits & Flat ON stack that is normalized to the Flat ON 99th percentile. &  & & \\
.flat\_on.json & A debugging output that reports the FWHM of the flat ON background distribution. &  & & \\
.wvlsol\_v1.fits & The night's wavelength solution determined from the OH sky emission lines. & Columns are x-pixel position on the detector and rows are echelle orders. The pixel values are the wavelength in microns. & & \\
joined\_flexure.csv & Reports the measured flexure along the x-axis in pixels that is used in the flexure correction for each band. & The first column is the frame number, the second column is the measured flexure in H-band, and the third column is the measured flexure in K-band & & \\
\enddata
\end{deluxetable*}

\section{IGRINS Pipeline Package (\texttt{IGRINS PLP})}\label{sec:PLP}
\subsection{PLP Data Processing}

The IGRINS Pipeline Package (\texttt{IGRINS PLP}) performs standard data reduction for IGRINS data. 
The \texttt{IGRINS PLP} removes cosmic rays (CRs), corrects for instrumental flexure,stacks the individual exposures for each slit position (A/B or ON/OFF), subtracts the stacked A/ON nods from the stacked B/OFF nods to remove sky emission features, removes detector readout pattern, applies the flat field, rectifies the individual echelle orders, calculates the wavelength solution, and extracts the per pixel flux and variance spectra.  
In this paper, we present version 3 (v3) of the \texttt{IGRINS PLP} \citep{Kaplan2024}, which includes several important updates over previous versions.
All the IGRINS data in RRISA have been reduced with \texttt{IGRINS PLP v3}.
While RRISA provides reduced IGRINS data products, the \texttt{IGRINS PLP} is open source and can be downloaded and run locally for custom data reductions. 
The
\texttt{IGRINS PLP} code and documentation are publicly available on the \href{https://github.com/igrins/plp}{IGRINS PLP GitHub} which includes detailed instructions for how to install and run.

Details of reduced data products can be found on the \href{https://github.com/igrins/plp/wiki/PLP-Data-Reduction-Products}{PLP Data Reduction Products} page of the \texttt{IGRINS PLP} wiki.
The final reduced data products are separated by band (H and K) and consist of extracted 1D and 2D spectra of the individual orders, with the x-axis representing the dispersion direction in detector pixels.  
The 2D spectra are straightened and rectified with the y-axis representing the spatial position along the IGRINS slit.  
Finally, the wavelength solution is given in vacuum wavelength in units of microns.

\subsection{Cosmic Ray and Pattern Noise Removal}

\begin{figure}
    \includegraphics[width=0.49\textwidth]{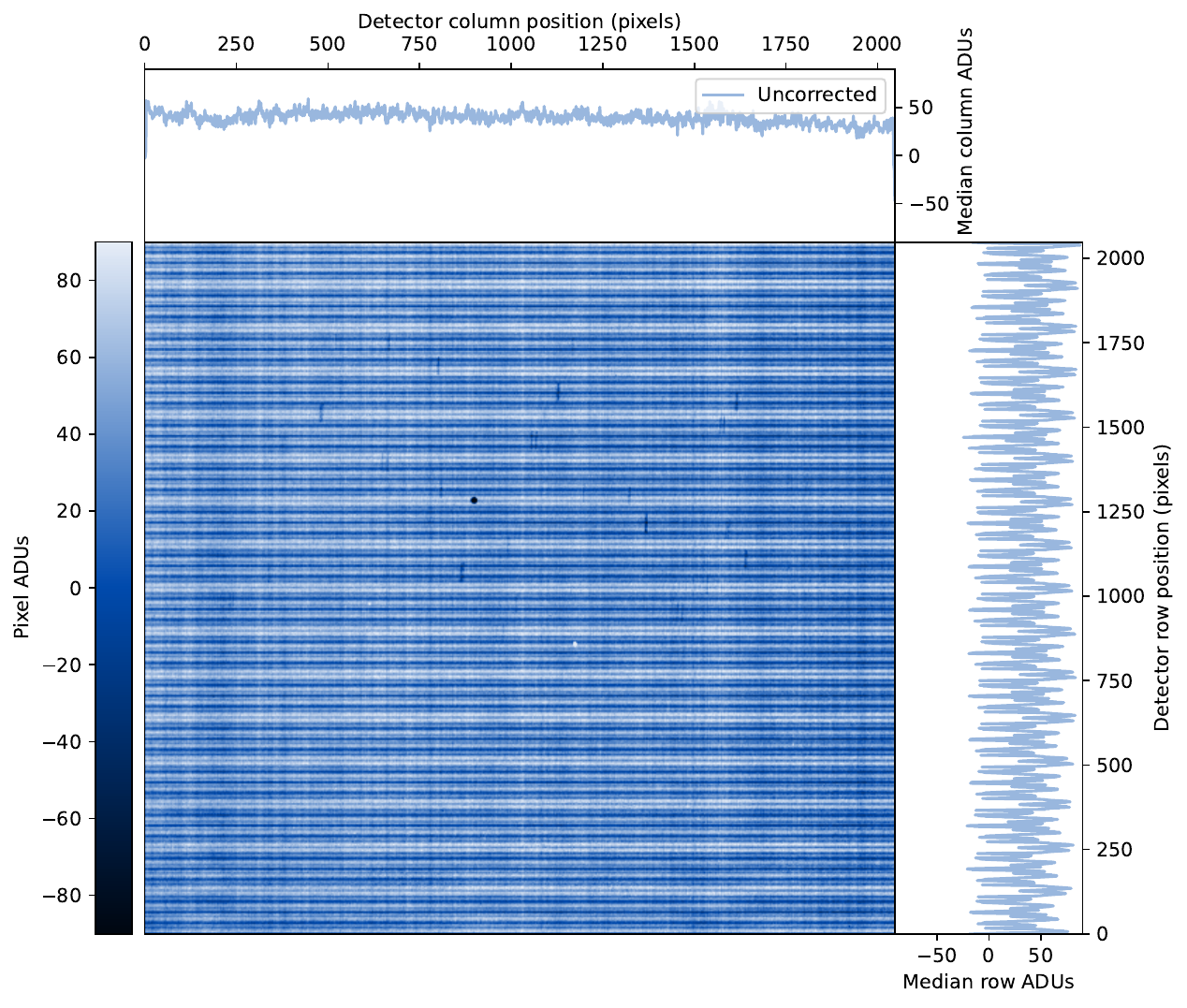}
    \includegraphics[width=0.49\textwidth]{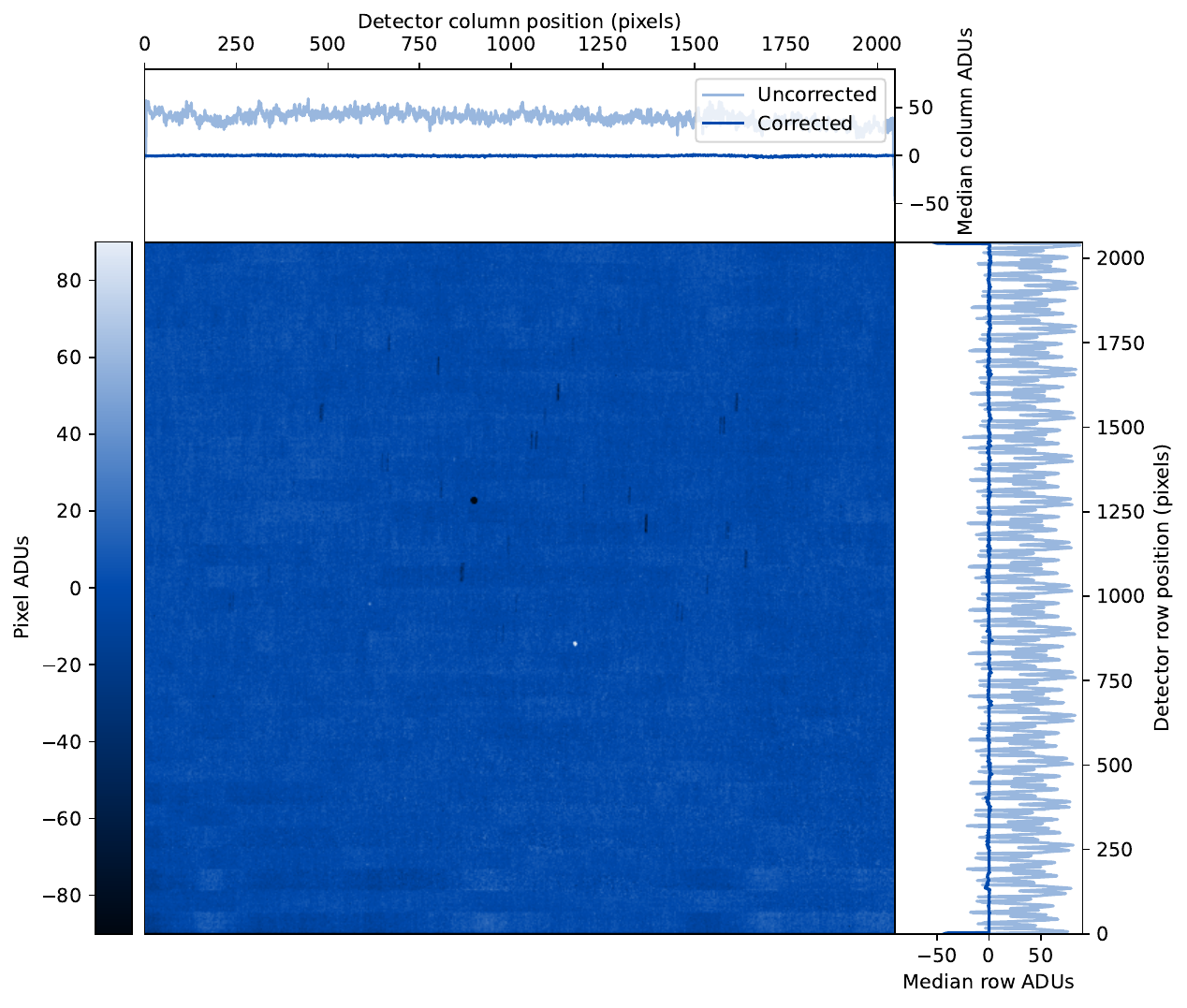}\\
    \includegraphics[width=0.49\textwidth]{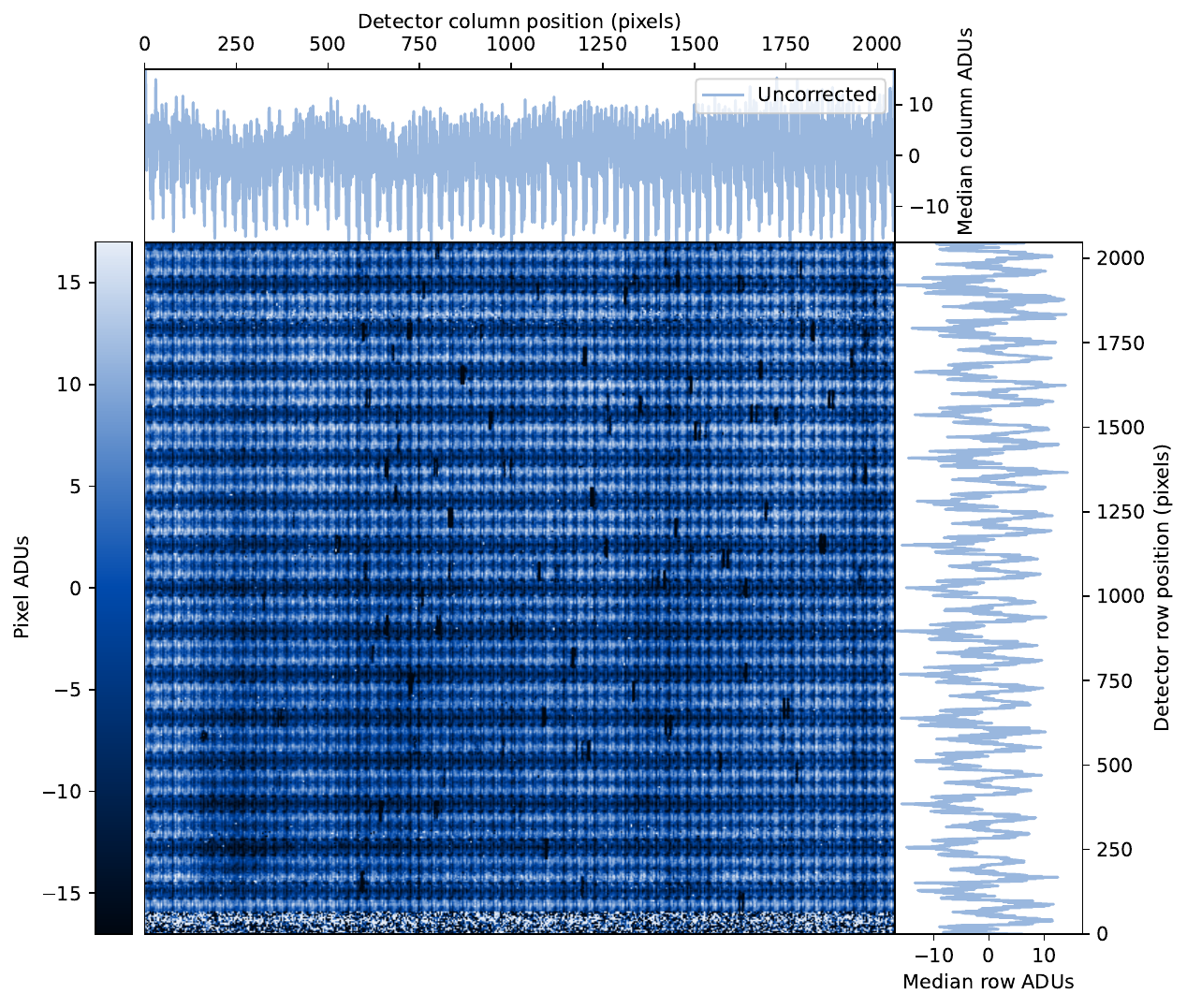}
    \includegraphics[width=0.49\textwidth]{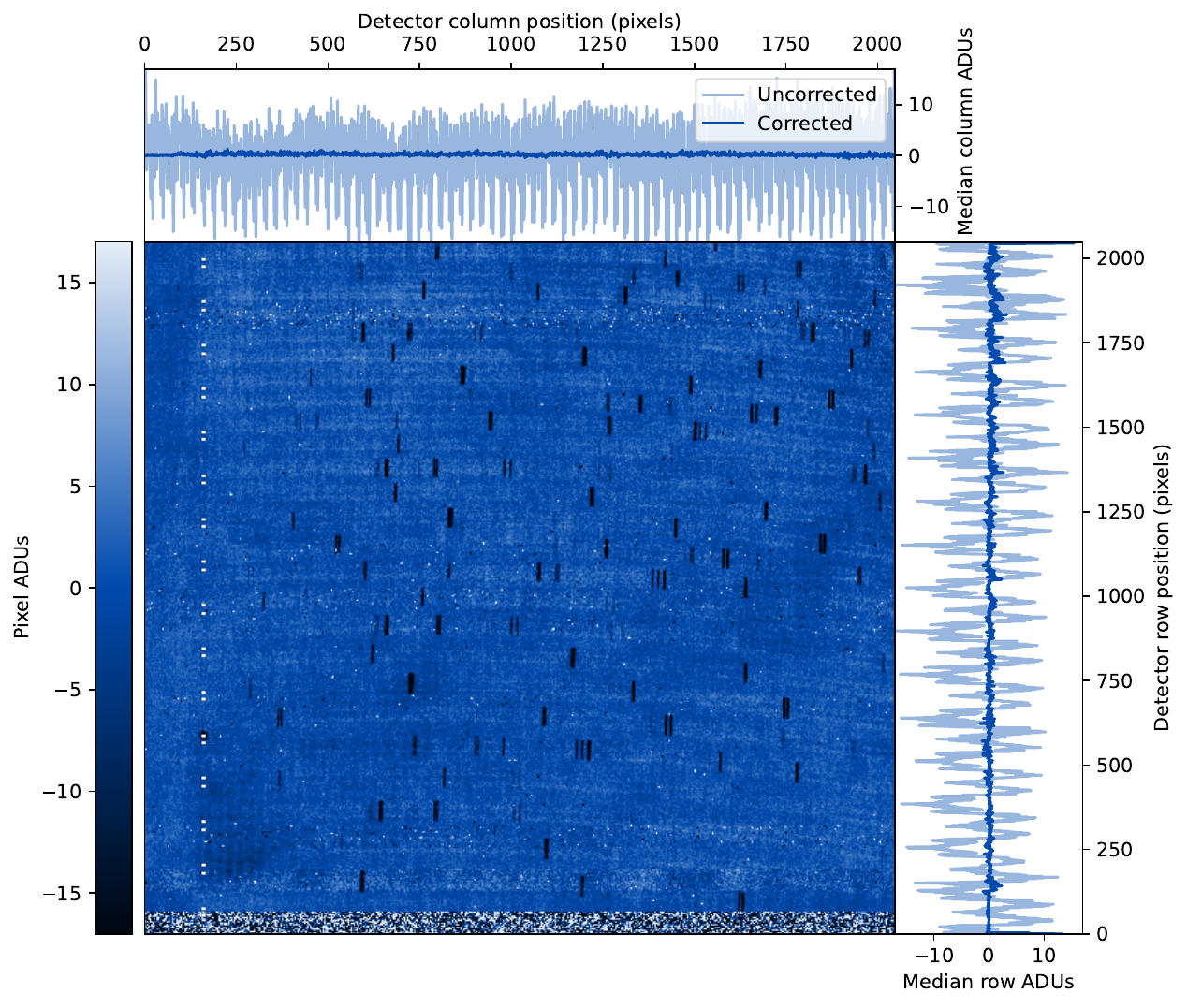}
    \caption{An example of IGRINS raw detector readout pattern (left column) and the correction (right column) for the 2D echellogram.
    The top and right line plots around each echellogram show the 1D median for the columns and rows in the 2D echellogram, respectively. The lighter blue in the 1D plots represents the median of the uncorrected data, and the darker blue shows the median of the corrected data. The top row shows the K-band detector from a night at McDonald Observatory in 2015. The bottom row shows the H-band detector from a night at Gemini South in 2024. The dark rectangles left over in the corrected 2D echellograms are residual OH sky emission lines.}
    \label{fig:pattern}
\end{figure}

\texttt{IGRINS PLP} v3 introduces CR correction to each individual raw exposure.
We identify CRs in individual exposures using \texttt{astroscrappy} \citep{mccully18}, an updated python version of \texttt{la-cosmic} \citep{dokkum2001}.  
To identify CRs, we tuned the parameters for H- and K-band exposures to minimize false identifications, such as the edges of OH sky emission lines.  
Since we identify CRs in the individual exposures and not stacked frames, \texttt{IGRINS PLP v3} will ``fill in'' the pixels in an exposure impacted by a CR. 
The impacted pixels are replaced by the median pixel value of all other exposures in the stack from the same nod (A or B, ON or OFF), each of which has been median smoothed to reduce the introduction of unwanted artifacts.  
The ``filling in'' method requires one exposure that overlaps the CR pixel without CR contamination and improves with more CR-free exposures.
If only a single exposure exists (i.e., only a single A-B nod was observed), the CR-affected pixels are filled in with pixels from \texttt{astroscrappy}'s built-in median smoothed clean array.
The final spectrum is much ``cleaner,'' and free of holes where the CRs impacted detector pixels.  For stellar sources, the \citet{horne86} optimal extraction procedure provides additional CR masking.

The raw IGRINS exposures show a detector readout pattern that regularly repeats.
The pattern is highly variable between the H- and K-band detectors, between exposures, and between telescopes.
An example of the typical detector readout pattern at HJST and LDT can be seen in the top left echellogram in Figure \ref{fig:pattern}.
The detector readout pattern worsened while IGRINS visited Gemini South due to increased electronic noise (see the bottom left echellogram in Figure \ref{fig:pattern}). 
The pattern repeats in regular intervals corresponding to the detector's 32 readout channels. 
The pipeline first removes the constant term (along the readout direction) using the values in the detector overscan reference columns. 
The pipeline further removes the pattern using information from the inter-order pixels. 
To achieve this, the pipeline mirrors and folds the stacked A-B frame (with sky-lines and thermal background removed by the subtraction) into 64-column segments to identify the repeating pattern and subtract it. 
Figure \ref{fig:pattern} shows example plots of the H- and K-band detectors before and after the readout pattern removal step in the pipeline.  The 1D plots of the median rows and columns show the dramatic improvement in the echellogram after removing the pattern.



\subsection{Flexure Correction}

\begin{figure}
    \center
    \includegraphics[width = \columnwidth]{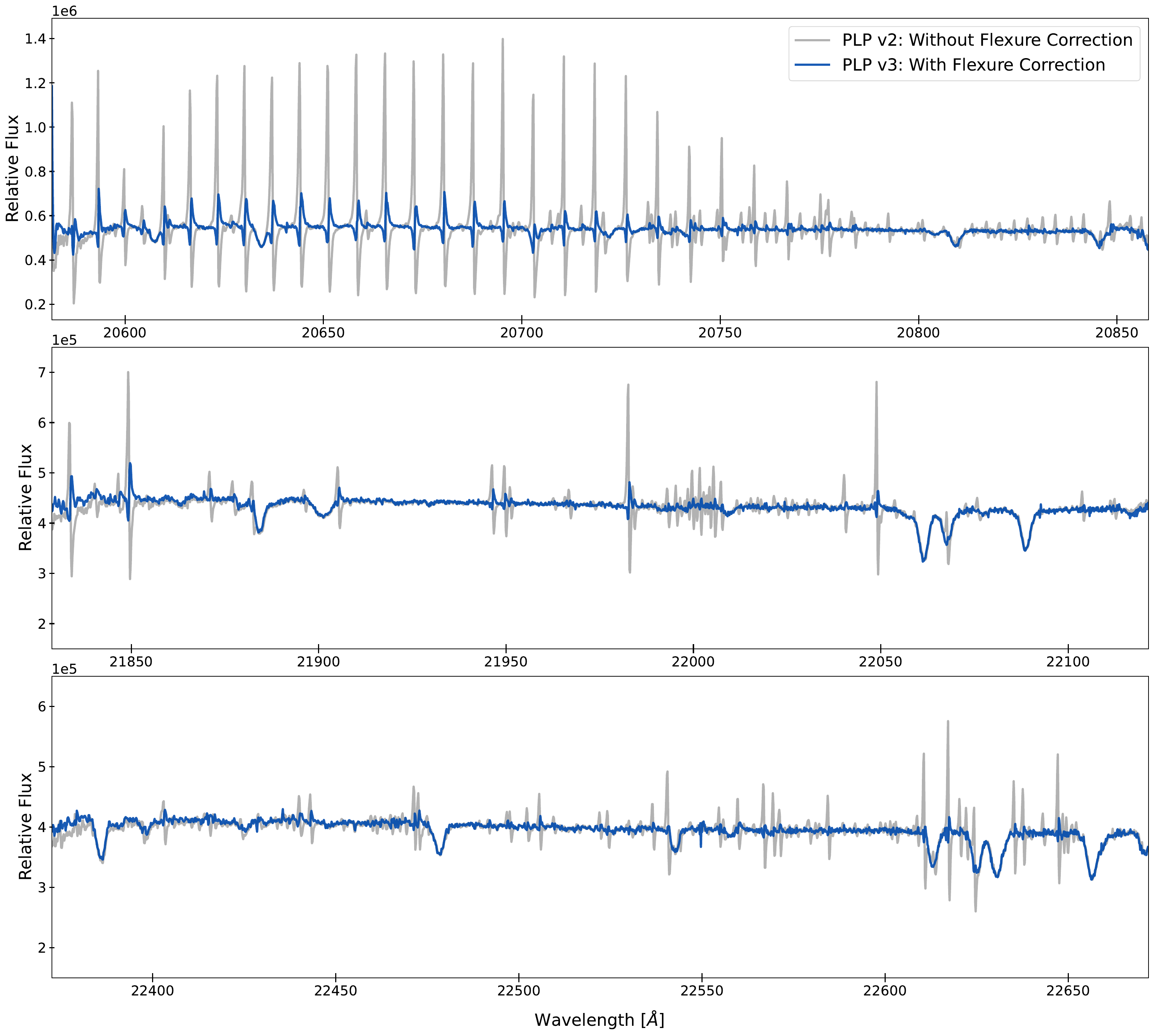}
    \caption{Three K-band orders for a target spectrum divided by an A0V standard star spectrum demonstrating how correcting the flexure improves the telluric line residuals. 
    The blue spectrum has the flexure correction applied, while the gray spectrum does not. The spikes in the 1D spectra are negative and positive residuals from deep telluric lines because the flexure has shifted the position of the spectrum on the detector between the target and standard star exposures. The remaining telluric residuals, seen in the \texttt{IGRINS PLP} v3 reduced data, are due to the airmass difference between the target and the telluric standard star. \emph{Note: This is data from 
    where the flexure shifts were larger than typical.}
    }
    \label{fig:flexure}
\end{figure}

Flexure, the shift in the position of instrument components due to changes in the gravity vector during observations, manifests in IGRINS spectra as small changes in the position of the echellogram on the detector. Flexure is a known feature of IGRINS data \citep{Park2014} and is discussed in some detail in \citet{Mace2016}.
A small pixel shift in the dispersion direction can lead to poor OH sky-line and background subtraction as well as poor telluric correction when dividing a science target by the standard star. Significant flexure in the dispersion direction can also degrade the spectral resolution when frames are stacked.
Flexure in the dispersion direction is typically low ($< 0.2$ pixels in the K-band) but can create shifts up to 1.5 and 0.15 pixels for the K- and H-band respectively.
The H-band flexure is typically an order of magnitude less than the K-band.


In \texttt{IGRINS PLP} v3, we introduce a frame-by-frame flexure correction using the OH sky-emission lines as a reference.  The flexure correction significantly reduces the misalignment of telluric absorption lines and sky emission lines between exposures on a target and between science and telluric standard star observations. 
Since the OH sky-emission lines are present in all exposures and are not dependent on the source continuum, flexure correction can be performed on nearly all science files.
A 300s SKY exposure taken each night for wavelength calibration is used as the flexure reference frame.  
For each raw exposure, the flexure is measured in the dispersion direction by cross-correlating the OH sky emission lines with the SKY reference frame.
Figure \ref{fig:flexure} illustrates how correcting the flexure between frames improves the telluric calibration and OH sky-line subtraction when dividing a target spectrum by a standard star. 

In detail, the flexure correction uses cross-correlation to find the pixel shift in the dispersion direction between the OH sky emission line positions between each exposure and the SKY reference frame.  
The echellogram in each exposure is smoothed and filtered to isolate the OH sky emission lines from other signals (e.g., stellar continuum) and then divided by the exposure time to normalize the OH line brightness to counts~s$^{-1}$.  
The echellogram is then masked with rectangular apertures to isolate bright OH lines and pixels outside the apertures are set to a value of zero.
The smoothed, normalized, and masked echellogram is sum collapsed along the y-axis, producing a one-dimensional array along the x-axis (the dispersion axis) of the OH sky-lines from all echelle orders. 
Then, the collapsed echellogram is enlarged $1000\times$ using linear interpolation.  
Cross-correlation is performed using a fast Fourier Transform by running \texttt{scipy.ndimage.fftconvolve} \citep{Virtanen2020} to measure the shift $\Delta x$ in pixels along the x-axis between the OH sky-line positions between the exposure and the SKY reference exposure.  
The cross-correlation function is limited to within $-2.5$ to $2.5$~pixels and
divided by a 4th-order polynomial to remove large-scale trends introduced by unwanted signals (e.g., bright stellar continuum).
Sub-pixel shifts ($\Delta x < 1$) are applied using linear interpolation.  
Two copies of the echellogram, with one shifted by a whole pixel, are averaged by weighting the amount of sub-pixel shift.  
If $\Delta x > 1$, the echellogram is rolled an integer number of pixels (almost always 1, very rarely 2) before the sub-pixel shift correction is applied for the remainder of $\Delta x$.  
This method minimizes smearing from linear interpolation to only the remainder of the flexure correction $< 1$ pixel.  
We note that the the sub-pixel correction can lead to a small loss in spectral resolution, although this loss in resolution is preferable to no flexure correction.  
If the flexure varies significantly over the target observation, then the flexure correction actually improves the spectral resolution of the final spectrum since it would otherwise be smeared in the dispersion direction by the varying flexure.

For short exposures ($< 20$ seconds), there will be less signal in the OH sky-lines per exposure, and flexure is determined using a slightly modified approach.  
Since the flexure does not change much over the course of short observations, one flexure correction is made for the entire observation.
The stack of all A/ON nods and B/OFF nods are summed, and the absolute value difference is determined as follows:  $A + B - |A-B|$.  
This scheme removes most of the signal from the target while leaving the signal from the OH sky-lines. 
The stack is treated as a single exposure, and the flexure correction is applied, as outlined above.

\section{Tools \& Tutorials}\label{sec:tools}
\subsection{Python Tools}
There are two open-source Python-based tools, \href{https://github.com/OttoStruve/muler}{\texttt{$\mu$ler}} \citep{Gully-Santiago2022} and \href{https://github.com/BrownDwarf/gollum}{\texttt{gollum}} \citep{Shankar2024}, that allow users to easily interact with reduced IGRINS data.
\texttt{$\mu$ler} helps users analyze and interact with reduced spectra from 3 NIR echelle spectrographs: IGRINS, the Habitable Planet Finder \citep{2012Mahadevan, 2014Mahadevan}, and Keck NIRSPEC \citep{1995McLean, 1998McLean, 2018Martin}.
There are many functions within \texttt{$\mu$ler} that help with additional data clean-up post-pipeline processing, including radial velocity shifting, removing flux outliers, stitching spectral orders, any mathematical manipulation of orders, measuring the equivalent widths of spectral features, and plotting (see Figure \ref{fig:flexure}) for both 1D and 2D IGRINS spectra.  
We encourage users of RRISA, especially those interacting with IGRINS data for the first time, to use \texttt{$\mu$ler} when handing IGRINS data using Python.

\texttt{gollum} is a more specialized tool that builds off of \texttt{$\mu$ler} and allows users to analyze and visualize synthetic spectral model grids used in estimating the astrophysical parameters of stars or brown dwarfs, like Sonora-Bobcat \citep{Marley2021, Marley2021b} or PHOENIX models \citep{1995Allard, 2013Husser}.
One unique feature of \texttt{gollum} is the interactive dashboard where users can tune synthetic model parameters like T$_{\textrm{eff}}$, $\log_{10}(\textrm{g})$, $[$Fe/H$]$, and v~$\sin(\textrm{i})$ with sliders to fit their data's spectral features.  \texttt{$\mu$ler} and \texttt{gollum} can be used together to compare stellar spectra to synthetic spectra from model grids, and the synthetic spectra can be used to match standard stars for empirical flux calibration and telluric correction. 

\subsection{Tutorials}
Once RRISA is downloaded, the information RRISA contains can be easily analyzed using a program like TOPCAT \citep{2005Taylor} or through Python.
We provide some tutorials for parsing information from RRISA using Python on the \href{https://igrinscontact.github.io/tutorials/}{Tutorials page} of the RRISA website. 
In particular, we provide examples for users to find the highest SNR spectra for each unique object, how to parse the list of SIMBAD aliases for each object in RRISA to find objects with specific names, and how to create a list of all of the targets (not including standard stars) IGRINS has ever observed.
Additionally, we show how users can download IGRINS reduced data products via Python using the Box links provided in RRISA, so users do not have to directly interface with Box.
Lastly, we supply two examples of how to plot reduced IGRINS spectra using \texttt{$\mu$ler}; one where we use the native \texttt{$\mu$ler} plotting function format and another that demonstrates how to customize \texttt{$\mu$ler} plotting with \texttt{matplotlib} \citep{Hunter2007}.

For users that are interested in customized reductions using the \texttt{IGRINS PLP} v3, we discuss how to download raw IGRINS data using RRISA, how to craft the recipe logs that are used to reduce data, and how to use the example \texttt{bash} script (\textit{run.sh}) to run the \texttt{IGRINS PLP} v3 recipes to reduce data.
For creating recipe files, we include solutions for common raw data issues, such as when nights are missing sky or flat frames.
We also provide details about each required column in a recipe log so users can successfully build recipe logs to reduce IGRINS data.
For users familiar with \texttt{IGRINS PLP}, these tutorials will be sufficient to reduce raw IGRINS data, but we also provide more detailed  \texttt{IGRINS PLP} v3 information on each step in the IGRINS data reduction process, explanations for each \texttt{IGRINS PLP} v3 output file, and more via the \href{https://github.com/igrins/plp/wiki}{\texttt{IGRINS PLP} v3 GitHub Wiki} for first-time users.

Our last tutorial section describes three cases where significant telluric residuals will persist in data, provides example data for each case, and suggests solutions for improving telluric residuals.
The first case occurs when an entire night of reduced data has persistent large telluric residuals; often, this results from a poor sky frame being used when reducing the data since a poor sky frame can lead to poor flexure correction and can be fixed by using a different sky frame during reduction.
The second case, the most common case, is where telluric residuals result due to a difference in airmass between the target and standard star.
Unfortunately, this case is the most difficult to resolve and requires using a telluric model to correct for the airmass differences.
The final case occurs when a frame where the target is not centered on the slit or the nod occurs in the wrong position on the slit is used in the data reduction, causing a telluric shift between the frames used for the reduction.
This final case can be difficult to troubleshoot as the change between a good and bad frame can be as subtle as a single pixel shift, but in the case where this problem is identified, removing the bad frame and re-reducing the data should resolve the issue.

\section{Conclusions}
We improved the IGRINS data reduction pipeline \texttt{IGRINS PLP} v3 by implementing a better CR and pattern noise removal procedure and a new flexure correction routine.
IGRINS data reduced using the new \texttt{IGRINS PLP} v3 often has significantly less telluric residual features, particularly in the K-band.
We introduce the Raw and Reduced IGRINS Spectral Archive, RRISA, which allows for \emph{free} access to raw and reduced IGRINS data for the first time in the 10 years IGRINS has been on-sky. 
RRISA inherits information from the SIMBAD, 2MASS, Gaia DR3, APOGEE2 DR17, and PASTEL catalogs for IGRINS targets with reduced data when available.
We discuss two Python-based tools compatible with IGRINS data, \texttt{$\mu$ler} and \texttt{gollum}, and outline the tutorials RRISA provides to support users.
Finally, we are planning an additional data release for RRISA in May 2025, when all of the data from Gemini South is released from its one-year proprietary period.
Interested users can visit the RRISA website at \url{igrinscontact.github.io}, download RRISA at \url{https://github.com/IGRINScontact/RRISA}, and download the newest version of the \texttt{IGRINS PLP} from \url{https://github.com/igrins/plp}.

\subsection{Acknowledge RRISA}
We ask that users of RRISA add the follwing acknowledgments to their papers:

\begin{displayquote}
The Raw and Reduced IGRINS Spectral Archive (RRISA) is maintained by the IGRINS Team with support from McDonald Observatory of the University of Texas at Austin and the US National Science Foundation under grant AST-1908892.

The Immersion Grating Infrared Spectrometer (IGRINS) was developed under a collaboration between the University of Texas at Austin and the Korea Astronomy and Space Science Institute (KASI) with the financial support of the US National Science Foundation under grants AST-1229522, AST-1702267 and AST-1908892, McDonald Observatory of the University of Texas at Austin, the Korean GMT Project of KASI, the Mt. Cuba Astronomical Foundation and Gemini Observatory.
\end{displayquote}

\section{Acknowledgements}
\noindent Facilities: 2.7m Harlan J. Smith Telescope at McDonald Observatory, 4.3m Lowell Discovery Telescope, and 8.1m Gemini South Telescope

\noindent Instruments: 2MASS \citep{Kleinmann1992, Stiening1995, Milligan1996, Skrutskie1997}, APOGEE \citep{Wilson2010, Wilson2012, Majewski2017}, Gaia \citep{Lindegren1994, Gilmore1998, Perryman2001, Gaia2016}, IGRINS \citep{Yuk2010, Park2014, Mace2016, Mace2018}

\noindent Software: \texttt{astropy} \citep{astropy:2013, astropy:2018, astropy:2022}, \texttt{astroquery} \citep{Ginsburg2019}, \texttt{barycorrpy} \citep{Kanodia2018b}, CDS \texttt{VizieR} \citep{Ochsenbein2000}, \texttt{IGRINS PLP} \citep{Sim2014, Lee2017, Kaplan2024}, \href{https://jupyter.org/}{\texttt{jupyter}}, \texttt{matplotlib} \citep{Hunter2007}, \texttt{numpy} \citep{harris2020array}, \texttt{pandas} \citep{McKinney2010, pandas2020}, \texttt{scipy} \citep{Virtanen2020}

\begin{center}
    \noindent\rule{3cm}{0.2pt}
\end{center}
This work used the Immersion Grating Infrared Spectrometer (IGRINS) was developed under a collaboration between the University of Texas at Austin and the Korea Astronomy and Space Science Institute (KASI) with the financial support of the US National Science Foundation under grants AST-1229522, AST-1702267 and AST-1908892, McDonald Observatory of the University of Texas at Austin, the Korean GMT Project of KASI, the Mt.\ Cuba Astronomical Foundation and Gemini Observatory.

This paper includes data taken at The McDonald Observatory of The University of Texas at Austin.

These results made use of the Lowell Discovery Telescope at Lowell Observatory. Lowell is a private, nonprofit institution dedicated to astrophysical research and public appreciation of astronomy and operates the LDT in partnership with Boston University, the University of Maryland, the University of Toledo, Northern Arizona University and Yale University. We are grateful for the generous donations of John and Ginger Giovale, the BF Foundation, and others, which made the IGRINS-LDT program possible. Additional funding for IGRINS at the LDT was provided by the Mt. Cuba Astronomical Foundation and the Orr Family Foundation.

Based on observations obtained at the international Gemini Observatory, a program of NSF NOIRLab, which is managed by the Association of Universities for Research in Astronomy (AURA) under a cooperative agreement with the U.S. National Science Foundation on behalf of the Gemini Observatory partnership: the U.S. National Science Foundation (United States), National Research Council (Canada), Agencia Nacional de Investigaci\'{o}n y Desarrollo (Chile), Ministerio de Ciencia, Tecnolog\'{i}a e Innovaci\'{o}n (Argentina), Minist\'{e}rio da Ci\^{e}ncia, Tecnologia, Inova\c{c}\~{o}es e Comunica\c{c}\~{o}es (Brazil), and Korea Astronomy and Space Science Institute (Republic of Korea).

This research made use of the cross-match service provided by CDS, Strasbourg.

This research has made use of the VizieR catalogue access tool, CDS, Strasbourg, France (\href{https://vizier.cds.unistra.fr/}{DOI:10.26093/cds/vizier}). The original description of the VizieR service was published in \href{https://doi.org/10.1051/aas:2000169}{2000, A\&AS 143, 23}.

This work presents results from the European Space Agency (ESA) space mission Gaia. Gaia data are being processed by the Gaia Data Processing and Analysis Consortium (DPAC). Funding for the DPAC is provided by national institutions, in particular the institutions participating in the Gaia MultiLateral Agreement (MLA). The Gaia mission website is \url{https://www.cosmos.esa.int/gaia}. The Gaia archive website is \url{https://archives.esac.esa.int/gaia}.

\bibliography{PASPsample631}{}
\bibliographystyle{aasjournal}



\end{document}